\documentclass[twocolumn,english,prl,twocolumn]{revtex4-1}
\usepackage{lmodern}
\usepackage{lmodern}
\usepackage[T1]{fontenc}
\usepackage[latin9]{inputenc}
\usepackage{graphicx}
\usepackage{esint}
\usepackage{upgreek}
\makeatletter

\usepackage{verbatim}
\usepackage{amstext}
\usepackage{babel}

\makeatother

\usepackage{babel}
\def\mr{\mathrm}
\def\mi{\mathit}
\begin{document}

\title{Seebeck coefficient of a single van der Waals junction in twisted bilayer graphene}

\author{Phanibhusan S. Mahapatra$^{1}$, Kingshuk Sarkar$^{1}$, H. R. Krishnamurthy$^{1}$,
Subroto Mukerjee$^{1}$ \& Arindam Ghosh$^{1}$}

\affiliation{$^{1}$Department of Physics, Indian Institute of Science, Bangalore
560 012, India. }



\begin{abstract}
When two planar atomic membranes are placed within the van der Waals distance, the charge and heat transport across the interface are coupled by the rules of momentum conservation and structural commensurability, lead to outstanding thermoelectric properties. Here we show that an effective 'inter-layer phonon drag' determines the Seebeck coefficient ($S$) across the van der Waals gap formed in twisted bilayer graphene (tBLG). The cross-plane thermovoltage which is nonmonotonic in both temperature and density, is generated through scattering of electrons by the out-of-plane layer breathing (ZO$^{'}$/ZA$_{2}$) phonon modes and differs dramatically from the expected Landauer-Buttiker formalism in conventional tunnel junctions. The Tunability of cross-plane seebeck effect in van der Waals junctions may be valuable in creating a new genre of versatile thermoelectric systems with layered solids

\end{abstract}
\maketitle

\vspace{20mm}
In spite of subnanometer separation of the van der Waals gap ($\sim0.5$~nm), the coupling of the two graphene layers in twisted bilayer graphene (tBLG) varies strongly with temperature ($T$), and the twist or misorientation angle $\theta$ between the hexagonal lattices of participating graphene layers \cite{koren2016coherent,perebeinos2012phonon,kim2013breakdown,dos2007graphene,boschetto2013real,cocemasov2013phonons,lui2015observation,luican2011single,ohta2012evidence}. At $T\ll\mi{\Theta}_{\mr{BG}}$, where $\mi{\Theta}_{\mr{BG}}$ is the Bloch-Gr${\ddot{{\rm u}}}$neisen temperature, the layers are coherently coupled either for $\theta\ll10^{\circ}$ with a renormalized Fermi velocity \cite{luican2011single,dos2007graphene}, or at specific values of $\theta$, such as $\theta=30^{\circ}\pm8.21^{\circ}$, when the hexagonal crystal structures become commensurate \cite{koren2016coherent}. For $\theta>10^{\circ}$ (and away from the `magic' angles), the layers are essentially decoupled at low $T$, but get effectively re-coupled at higher $T$ ($>\mi{\Theta}_{\mr{BG}}$), when the interlayer phonons drive cross-plane electrical transport through strong electron-phonon scattering \cite{perebeinos2012phonon,kim2013breakdown}. These phonons are also expected to determine thermal and thermoelectric transport across the interface \cite{ding2016interfacial,zhang2015thermal,sadeghi2016cross,chen2015thermoelectric,hung2014enhanced,juang2017graphene}. In fact, since the in-plane transverse and longitudinal phonons are effectively filtered out from contributing to cross-plane transport because they do not substantially alter the tunneling matrix elements, theoretical calculations predict enhanced cross-plane thermoelectric properties in van der Waals heterojunctions, including high \textit{ZT} factors at room temperature \cite{sadeghi2016cross}. However, although the impact of interlayer coherence and electron-phonon interaction on electrical conductance has been studied in detail \cite{koren2016coherent,kim2013breakdown}, their relevance to the thermal and thermoelectric properties of tBLG remains unexplored.

We assembled the tBLG devices with layer-by-layer mechanical transfer method, which is common in van der Waals epitaxy \cite{roy2013graphene,karnatak2016current,zomer2011transfer}. Three devices were constructed which show very similar behavior, and we present the results from one of the devices here. The device consists of two graphene layers oriented in a ``cross'' configuration (inset of Fig.~1a and an optical micrograph in Fig.~1b), and entirely encapsulated within two layers of hexagonal boron nitride (hBN). The carrier mobilities in the upper and lower layers are $\approx25000$~cm$^{2}$V$^{-1}$s$^{-1}$ and $\approx60000$~cm$^{2}$V$^{-1}$s$^{-1}$, at room temperature, respectively. More detail on the fabrication process can be found in Methods. The doping density can be varied with the global back gate (heavily doped silicon substrate), as well as locally at the overlap region by using a lithographically defined top gate. Multiple contacts on all sides of the overlap region allow a four-probe measurement of both in-plane and cross-plane transport.

\begin{figure*}[t]
\includegraphics[clip,width=18cm]{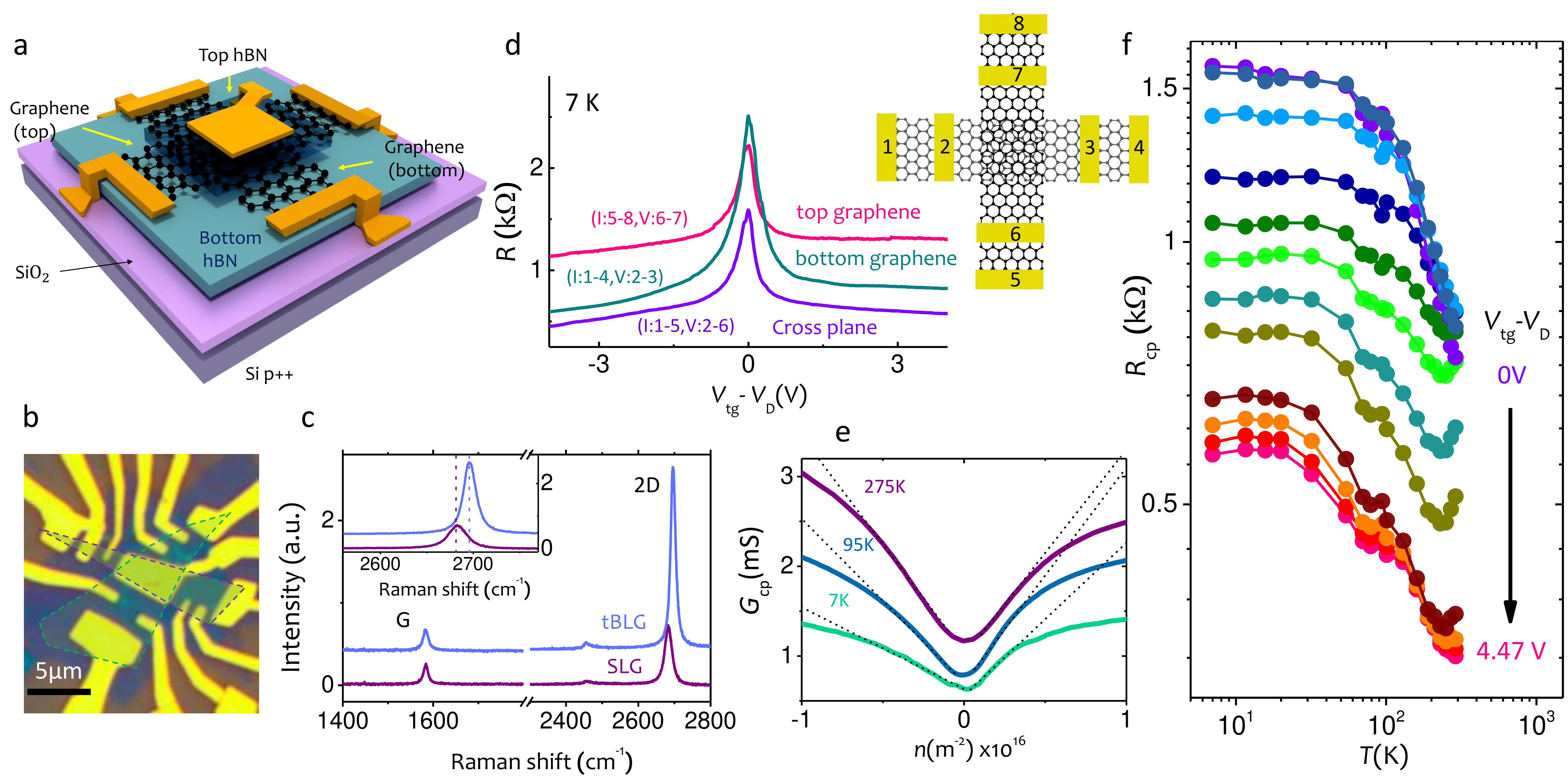}
  \caption{\textbf{ Device structure, Characterization and electrical transport}. (a) Schematic of the hBN-encapsulated twisted-bilayer graphene (tBLG) device geometry including the contact pads used for the cross-plane thermoelectricity measurement. (b) Top-view optical image of the device. (c) Comparison of Raman spectra from the tBLG and monolayer graphene regions. Inset shows the blueshift of the 2D peak from the monolayer graphene region relative to that at the tBLG region. (d) In-plane and cross-plane electrical resistances as functions of the top gate voltage. The corresponding contact configuration is shown in the inset. (e) Measured cross-plane electrical conductance ($G_{\mr{cp}}$) as a function of number density for three different temperatures. The dotted lines identify the linear regions in the variation of $G_{\mr{cp}}$. (f) Temperature dependence of cross-plane the electrical resistance ($R_{\mr{cp}}$) for different gate voltages.}
 \label{fgr:1}
\end{figure*}

The difference in the Raman spectra from the overlap and the bare graphene regions in Fig.~1c, allows an estimation of the twist angle $\theta$ \cite{kim2012raman,boschetto2013real,havener2012angle,he2013observation,lui2015observation,campos2013raman}. For optical transitions in the parallel band model \cite{havener2012angle,kim2012raman}, the blue shift, intensity enhancement and width reduction of Raman peaks is attributed to van Hove singularities in the presence of weak interlayer interaction. The observed blue shift of $\approx13$~cm$^{-1}$ in the 2D peak position (inset of Fig.~1c) suggests $\theta\approx13^{\circ}$, which is supported by the 2D peak width reduction and G band intensity as well (Fig.~S1 of supplementary information).

Fig.~1d shows the effect of top gate voltage ($V_{\mr{tg}}$) on both in-plane and cross-plane electrical resistance, while the back gate voltage $V_{\mr{bg}}$ is held fixed at $=-35$~V to minimize the contribution of series resistance. The in-plane bipolar transfer characteristics are expected in graphene, while similar observation in the cross-plane transport can arise from two processes: (1) density-of-state-dependent incoherent tunneling across the van der Waals gap with interlayer conductance $G_{\mr{cp}}=R_{\mr{cp}}^{-1}\propto D_{1}(E_{\mr{F}})D_{2}(E_{\mr{F}})\mi{\Gamma}$ \cite{kim2013breakdown,bistritzer2010transport}, or (2) interlayer conduction limited by electron-phonon scattering \cite{perebeinos2012phonon} with $G_{\mr{cp}}\propto n_{\mr{ph}}E_{\mr{F}}^{2}/\mi{\Omega}_{\mr{ph}}$, where $R_{\mr{cp}}$, $D_{\mr{i}}(E)$, $\mi{\Gamma}$, $E_{\mr{F}}$ and $n_{\mr{ph}}$ are the cross-plane electrical resistance, energy-dependent density-of-states of the $i$th layer, interlayer transmission probability, Fermi energy of the graphene layers, and the thermal population of the out-of-plane beating phonon mode of energy $\mi{\Omega}_{\mr{ph}}$, respectively. Although both mechanisms lead to $G_{\mr{cp}}\propto n$, in agreement with the observations for small $n$ (Fig.~1e), they differ in their temperature dependences. As shown in Fig.~1f, the low $T$ cross-plane transport is $T$-independent, which is consistent with incoherent quantum tunneling, whereas $R_{\mr{cp}}$ decreases sharply for $T>70$~K, suggesting the onset of phonon-driven electrical conduction as the thermal population of interlayer phonons increases with increasing $T$. Since both mechanisms depend on $n$ in a similar manner,  the crossover temperature scale ($\sim70$~K) varies weakly with doping, but serves as an indicator of the energy scale of interlayer phonons \cite{cocemasov2013phonons}. The slight asymmetry at low temperature ($7$~K) between the electron and the hole doped regimes is probably a series resistance effect outside the overlap region, because the two graphene layers exhibit different mobilities.

To measure the Seebeck effect across the van der Waals junction, we employ local Joule heating of one of the graphene layers, which establishes a  interlayer temperature difference $\Delta T$, while measuring the resulting thermal voltage generated between the layers (See schematic in Fig.~2a, and Methods). For a sinusoidal heating current $I_{\mr{h}}(\omega)$ of frequency $\omega$, the thermal component is obtained from the second harmonic ($V_{2\omega}$) of the cross-plane voltage \cite{zuev2009thermoelectric,goswami2009highly}. At fixed $T$, $V_{2\omega}$ varies with doping in a qualitatively similar manner as that observed for in-plane Seebeck effect in graphene \cite{zuev2009thermoelectric,checkelsky2009thermopower} (Fig.~2b), which is antisymmetric across the Dirac point, with the sign of $V_{2\omega}$
representing that of the majority carriers. The thermal origin of $V_{2\omega}$ can be readily verified from the scaling of $V_{2\omega,\mr{rms}}\propto I_{\mr{h,rms}}^{2}$
over the entire range of $V_{\mr{tg}}$ (Fig.~2c). The interlayer temperature gradient $\Delta T$ is determined entirely by $I_{\mr{h}}$ and $\Delta T\propto I_{\mr{h,rms}}^{2}$ (Inset of Fig.~2c) confirms that the thermoelectric power $S=V_{2\omega}/\Delta T$ is independent of $I_{\mr{h}}$ within the range of heating current ($\leq4~\mr{\mu}$A) of our experiment (see Methods and supplementary information for details on the temperature calibration).

\begin{figure*}[t]
\includegraphics[clip,width=18cm]{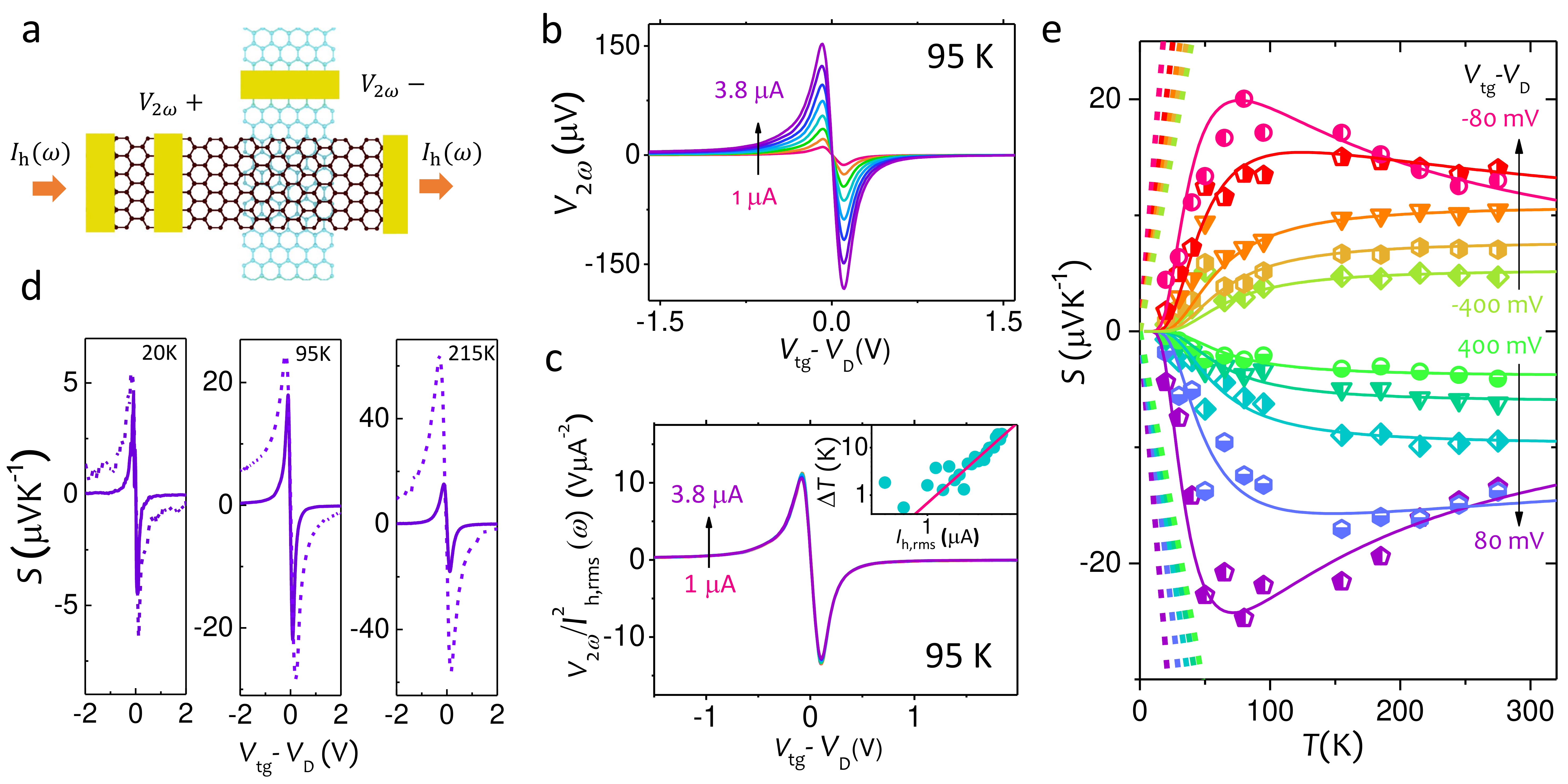}
  \caption{\textbf{Thermoelectricity in twisted bilayer graphene}.(a) Heating and measurement scheme for the  evaluation of thermoelectric parameters. (b) $2^{\mr{nd}}$ harmonic voltage $V_{2\omega}$ between two graphene layers as a function of the top gate voltage for different heating currents at a fixed temperature ($95$ K). (c) $V_{2\omega}$ scaled with $I_{\mr{h,rms}}^{2}$. The inset shows that the interlayer temperature difference $\Delta T$ is proportional to $I_{\mr{h,rms}}^{2}$. (d) Comparison between the measured Seebeck coefficient (solid line) and that calculated (dashed line) from the  semiclassical Mott relation (Eq.~\ref{eq1}) for three different temperatures. (e) The temperature dependence of the Seebeck coefficient for different $|V_{\mr{tg}}-V_{\mr{D}}|$. The dashed lines show the corresponding values of $S$ calculated from the Landauer-Buttiker formalism . The solid lines show fits from Eq.~\ref{eq2} for different doping values.}
\label{fgr:2}
\end{figure*}

We first compared the dependence of $S$ on $V_{\mr{tg}}$ to that expected from the semiclassical Mott relation \cite{jonson1980mott},

\begin{equation}
S_{\mr{Mott}}=-\frac{\pi^{2}k_{\mr{B}}^{2}T}{3|e|}\frac{d\ln{G_{\mr{cp}}}}{dE}|_{E=E_{\mr{F}}}
\label{eq1}
\end{equation}

\noindent at three different values of $T$. Evaluating the right hand side of Eq.~\ref{eq1} using the parallel plate capacitor model and linear dispersion of graphene, shows that the experimentally observed $S$ bears only a qualitative similarity to $S_{\mr{Mott}}$ (shown as dotted lines in Fig.~2d), and decreases far more rapidly with increasing $|V_{\mr{tg}}-V_{\mr{D}}|$, than that expected from Eq.~\ref{eq1}. A violation of the Mott relation \cite{goswami2009highly,buhmann2013thermoelectric,behnia2004thermoelectricity} is possible in the presence of strong electron-electron interaction, or localized magnetic moments, but this cannot explain the enhanced suppression of $S$ from the Mott relation at higher doping (\textit{i.e.} larger $|V_{\mr{tg}}-V_{\mr{D}}|$), where the interaction effects are expected to be minimal. Fig.~2d also shows that the suppression of the observed $S$ from the Mott relation becomes stronger at higher $T$, suggesting a likely role of interlayer phonons. To explore this further, the $T$-dependence of $S$ at different $|V_{\mr{tg}}-V_{\mr{D}}|$ is shown in Fig.~2e. The maximum $S$ of $\approx20-25~\mr{\mu}$VK$^{-1}$ occurs at $(V_{\mr{tg}}-V_{\mr{D}})\approx80$~mV, corresponding to $n\approx1\times10^{11}$~cm$^{-2}$, at $T\approx70$~K. The generic behavior of $S$ seems to indicate a characteristic temperature scale, which increases with $|V_{\mr{tg}}-V_{\mr{D}}|$, beyond which $S$ either weakly decreases (at low doping) or saturates to a finite value asymptotically (at high doping).

Modelling the tBLG as a tunnel junction with electronically decoupled graphene electrodes, the thermoelectric parameters can be directly computed with the Landauer-Buttiker formalism (See Fig.~3a and Methods for calculation details). At high doping, \textit{i.e.} $E_{\mr{F}}/k_{\mr{B}}T\gg1$, the calculated (electron/hole symmetric) thermopower decreases along a universal trace as a function of $E_{\mr{F}}/k_{\mr{B}}T$ (Fig.~3b). However, as $|n|\rightarrow0$, the inhomogeneity in the charge distribution, represented by a finite broadening of the dispersion relations in the layers, causes $S$ to vary in an inhomogeneity and temperature specific manner. A comparison in Fig.~3c reveals that the computed magnitude of $S$ from the Landauer-Buttiker formalism is $\sim1-3$ orders of magnitude larger than the experimentally observed $S$, especially at large $|n|$, where the experimental $S$ decays much faster than that expected theoretically (dashed lines in Fig.~3c).

\begin{figure*}[t]
\includegraphics[clip,width=18cm]{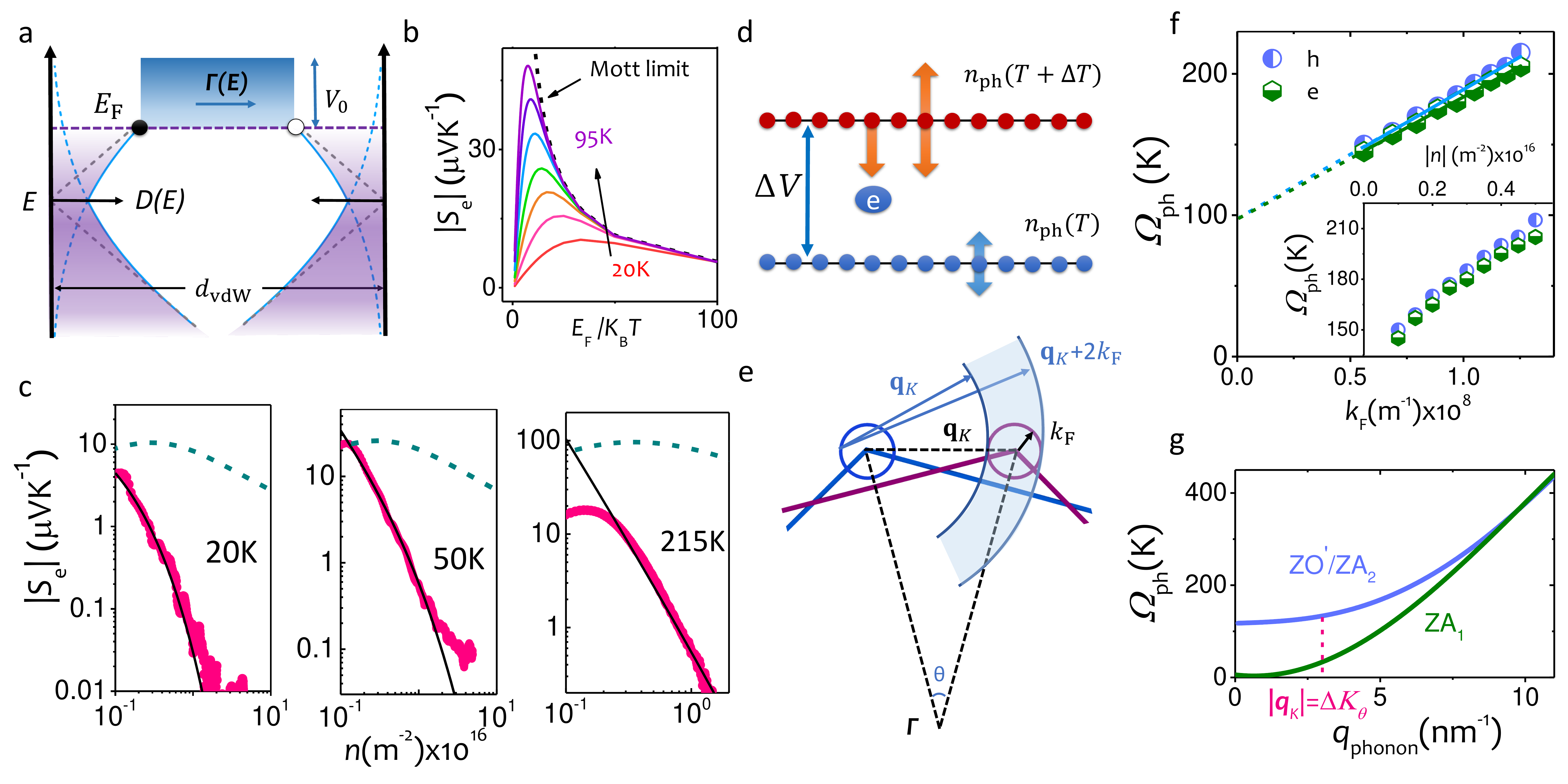}
  \caption{\textbf{ Microscopic mechanism of thermoelectric transport and layer breathing phonon modes.} (a) Schematic of the metal-insulator-metal junction model with charge-puddle broadened densities of states. (b) Calculated $S$ as a function of $E_{\mr{F}}/k_{\mr{B}}T$ for different temperatures (20K, 30K, 40K, 50K, 65K, 80K and 95K) with a broadening constant $\gamma =55$~meV. (c) Doping (electron) dependence of the experimentally measured $S$ (pink) at three temperatures. Dashed lines show the corresponding $S$ calculated from the metal-insulator-metal junction model. Solid lines (black) show the fits from Eq.~\ref{eq2}. (d) Schematic of the temperature gradient driven imbalance of phonon populations in the two graphene layers. (e) Schematic to explain (in-plane) momentum conserving eletron scattering from one graphene layer to another by layer breathing phonon modes. (f) Energy of the interlayer phonon mode  obtained from fitting Eq.~\ref{eq2} to $S-T$ data, as a function of Fermi wave vector $k_{\mr{F}}$ for both electron and hole doping. The intercept of the linear fit determines the energy of the zone centre phonon mode while the slope yields the momentum of the phonon mode corresponding to the momentum mismatch between the two rotated Dirac cones. The inset shows the interlayer phonon energy as a function of the number density. (g) Calculated energy dispersion of low energy interlayer optical and acoustic phonon modes \cite{perebeinos2012phonon}. The vertical dashed line evaluates the phonon branch energy and the corresponding phonon momentum, which compares well with that obtained from the analysis in (f).}
\label{fgr:3}
\end{figure*}

The observation of $S\rightarrow0$ at low temperatures where the electrical conductance occurs through incoherent tunneling (Fig.~2e), suggests that thermoelectric transport in tBLG is likely driven by the electron-phonon coupling. A phenomenological description involves charge imbalance across the layers induced by the imbalance in the thermal population ($n_{\mr{ph}}$) of phonons so that $\Delta V\propto|dn_{\mr{ph}}/dT|\times\Delta T$, where $\Delta V$ is the interlayer potential difference (schematic in Fig.~3d). Hence,

\begin{equation}
S=\frac{\Delta V}{\Delta T}=A(E_{\mr{F}})|\frac{dn_{\mr{ph}}}{dT}|=A(E_{\mr{F}})\frac{\mi{\Omega}_{\mr{ph}}}{T^{2}}\frac{e^{\mi{\Omega}_{\mr{ph}}/T}}{(e^{\mi{\Omega}_{\mr{ph}}/T}-1)^{2}}
\label{eq2}
\end{equation}

\noindent where $\mi{\Omega}_{\mr{ph}}(\mathbf{q}_{K},k_{\mr{F}})$ is the energy of the interlayer phonon mode that conserves momentum during the transfer of a charge from the Fermi surface of one graphene layer to that of the other ($\mathbf{q}_{K}$ and $k_{\mr{F}}=\sqrt{\pi n}$ represent the vector in the reciprocal space connecting the Dirac points of the rotated Brillouine zones, and the Fermi wave vector, respectively). The prefactor $A(E_{\mr{F}})$ embodies the electron-phonon coupling. Eq.~\ref{eq2} provides excellent fit to the observed $T$-dependence of $S$, shown by the solid lines in Fig.~2e, and allows us to estimate the $\mi{\Omega}_{\mr{ph}}$ which sets the characteristics scale in the $T$ dependence as a function of $n$ (inset of Fig.~3f). (In fitting the $S-T$ data for $|V_{\mr{tg}}-V_{\mr{D}}|<160$~mV, a weak $T$-dependence of $A\propto T^{-\beta}$, where $\beta\approx0.3-0.6$ was assumed.)

The observed magnitude of $\mi{\Omega}_{\mr{ph}}\sim150-200$~K, suggests that the low-energy ZO$^{'}$/ZA$_{2}$ layer breathing phonon modes \cite{cocemasov2013phonons,campos2013raman} in tBLG, which determines the interlayer electrical conductance \cite{perebeinos2012phonon,kim2013breakdown} are also primarily responsible for thermoelectric transport across the van der Waals gap. However, to cross-verify this quantitatively, we assume the phenomenological phonon dispersion of the out-of-plane breathing modes \cite{perebeinos2012phonon}, $\mi{\Omega}_{\mr{ph}}=\sqrt{\mi{\Omega}_{\Gamma}^{2}+ \frac{\kappa}{\rho} (\frac{h}{k_\mr{B}})^2(|\mathbf{q}_{K}|+k_{\mr{F}})^{4}}$, where the momentum conservation requires phonons with average momentum $\approx|\mathbf{q}_{K}|+k_{\mr{F}}$ to transfer charge between all points of the two Fermi surfaces ( schematic of Fig.~3e) (here, $\mi{\Omega}_{\Gamma}$, $\kappa$ and $\rho$ are the zone center phonon energy, bending stiffness and areal mass density of graphene, respectively). Since $|\mathbf{q}_{K}|\gg k_{\mr{F}}$, one expects $\mi{\Omega}_{\mr{ph}}$ to have a linear dependence on $k_{\mr{F}}$ , as indeed observed in Fig.~3f, and the slope of the linear dependence yields $|\mathbf{q}_{K}|\approx4$~nm$^{-1}$. This corresponds to a misorientation angle of $\theta\approx13^{\circ}$, in excellent agreement with the estimate of $\theta$ from Raman measurements (Fig.~1c). The intercept, $\mi{\Omega}_{\mr{ph}}(|n|=0)\approx101$~K, also agrees well with the ZO$^{'}$/ZA$_{2}$ phonon branch energy ($\sim115$~K) for $|\mathbf{q}_{K}|\approx4$~nm$^{-1}$ \cite{perebeinos2012phonon,cocemasov2013phonons,campos2013raman}, as shown by the vertical dashed line in Fig.~3g. Eq.~\ref{eq2} also captures the density dependence of $S$ shown in Fig.~3c if we choose a prefactor $A\sim n^{-\alpha}$ (solid lines Fig.~3c), although $\alpha$ is found to be $T$-dependent, varying from $\approx0$ at low $T$
to $\approx2$ at $T=215$~K.

\begin{figure*}[t]
\includegraphics[clip,width=9cm]{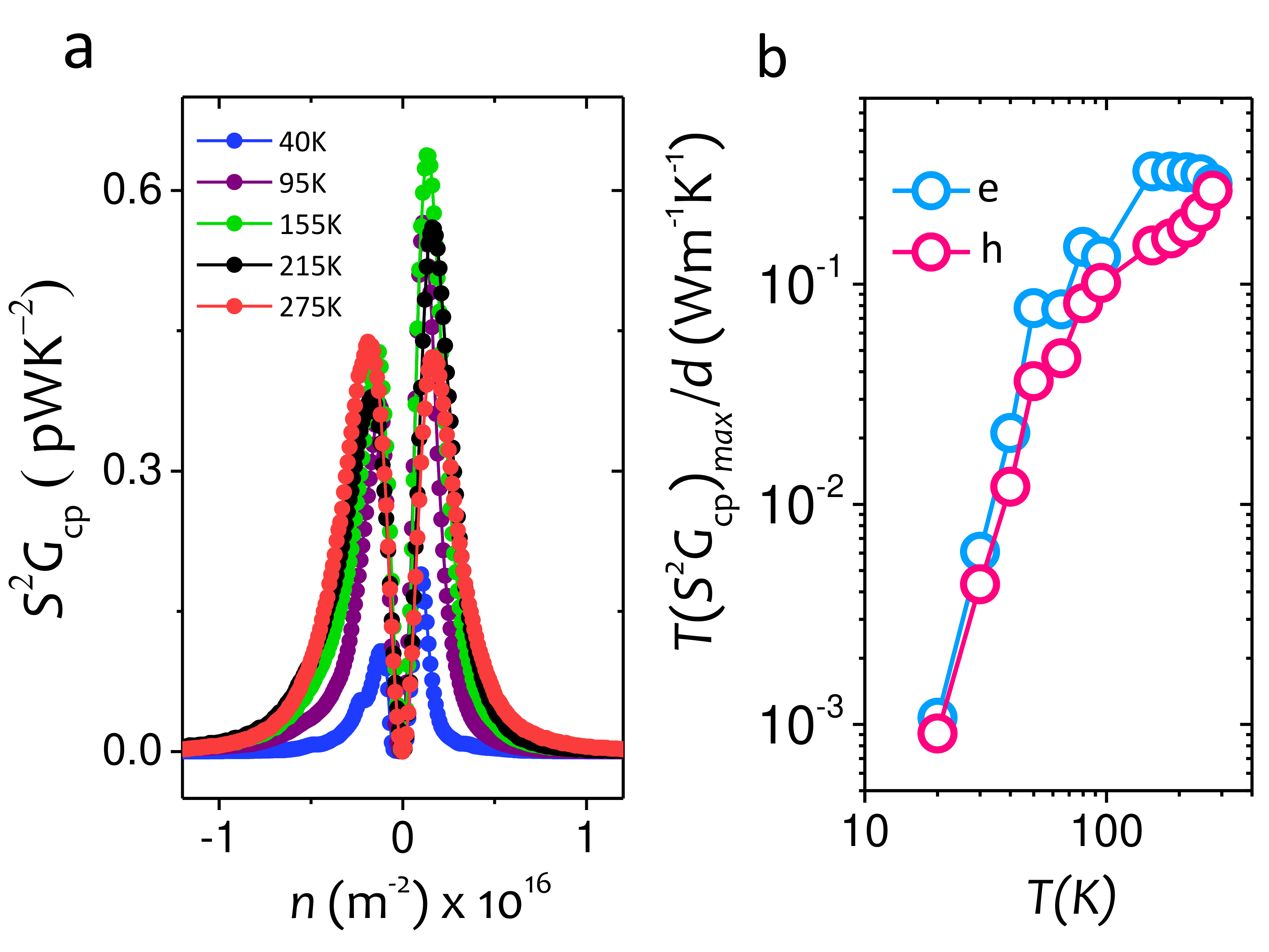}
  \caption{\textbf{Estimation of power factor in twisted bilayer graphene}. (a) Density dependence of cross-plane thermoelectric power factor calculated from the measured $S$ and $G_{\mr{cp}}$ at different temperatures. (b) Temperature dependence of peak value of PFT ($=TS^{2}G_{\mr{cp}}/d$).}
\label{fgr:4}
\end{figure*}

The deviation of $S$ from the Mott relation (Eq.~\ref{eq1}) is a key result of our experiments which demands further elaboration. The Mott relation is expected to hold even in the presence of phonon-driven tunneling, when the tunneling is isotropic (i.e. its amplitude depends only on the magnitude and not the direction of the momentum vector of the electron) and when the phonons are in thermal equilibrium. A non-equilibrium distribution of phonons can give rise to `cross-plane phonon drag' (due to a transport of heat from one layer to another by the layer breathing modes in each) resulting in a violation of the Mott formula. Furthermore, and rather unusually, the measured thermopower is always smaller than the value obtained from the Mott formula (Fig.~2d). A drag component in the presence of umklapp scattering could, in principle reduce $S$ below the regular tunneling component given by Landauer-Buttiker formula and hence deviate from the Mott relation \cite{bailyn1967phonon}. However, this is inconsistent with the observed dominance and the sign of the drag component in our devices. Hence, the violation of the Mott relation we observe is most likely due to a strong suppression of the regular tunneling component from the Mott value due to anisotropic tunneling arising from the twist between the layers.

Finally, we have calculated the cross-plane thermoelectric power-factor ($S^{2}G_{\mr{cp}}$), by combining the experimentally observed magnitudes of $S$ and $G_{\mr{cp}}$. Fig.~4a presents the nearly electron-hole symmetric power factor with a maximum value of $\approx0.5$~pWK$^{-2}$. The maximum in power-factor, observed at $|n|\approx0.1\times10^{11}$~cm$^{-2}$ in our case, is determined by the interplay of the increase of $G_{\mr{cp}}$ and decrease of $S$ with $|n|$, and can further be improved in higher mobility devices, where the onset of inhomogeneous transport occurs at lower $|n|$. The maximum effective PFT $=TS^{2}G_{\mr{cp}}/d$, where $d\approx0.4$ nm is the van der Waals distance, increases with temperature,  and can be as high as $\approx0.3$ Wm$^{-1}$K$^{-1}$ at room temperature (Fig.~4b). While this is about an order of magnitude smaller than in-plane PFT of high-mobility graphene \cite{duan2016high} and transition metal dichalcogenides layers \cite{yoshida2016gate,hippalgaonkar2015record}, the phonon-filtering in cross-plane thermal transport may lead to lower thermal conductance, and hence a high figure-of-merit thermoelectric system \cite{pop2012thermal,alofi2013thermal}.

In conclusion, we have measured, for the first time, the thermoelectric properties across the van der Waals gap formed in twisted bilayer graphene. We demonstrated that the cross-plane thermoelectric transport is driven by the scattering of electrons and interlayer layer breathing phonon modes, which thus represents a unique ``phonon drag'' effect across atomic distances. Although the deviation from the Mott relations needs further understanding, we believe that clever engineering of van der Waals heterostructures, for example, inclusion of intermediate atomic layers from layered solids acting as phonon filters \cite{chen2015thermoelectric}, may lead to exceptional cross-plane thermoelectric properties.

\section{Methods}

\subsection{Fabrication of hBN encapsulated tBLG devices}

All devices in this work were fabricated using layer
by layer mechanical transfer method where the overlap region of the two graphene layers, which forms the tBLG system, is encapsulated within two hBN layers to prevent surface contamination and minimize substrate and lithography effects. The top hBN layer also acts as the dielectric for the lithographically defined metal top gate. The electrical contacts to the individual layers were patterned using electron beam lithography (EBL). The electrical leads were mostly formed by etching through the hBN encapsulation, followed by a metallization step for edge-contacting the graphene layers. To achieve this, patterned contacts were exposed to reactive ion etching (RIE) to etch the top hBN. The metal deposition (5nm Cr/50nm Au) was then done by thermal evaporation technique to make electrical contact with the one dimensional edges of the single layer graphene channels.

\subsection{Calculation of the temperature gradient across the van der Waals gap.}

The temperature difference $\Delta T$ between the two monolayers is obtained by calibrating the individual in-plane resistances of the two monolayers as functions of both temperature and heating current at the charge neutrality point. The scaling of $V_{2\omega,\mr{rms}}\propto I_{\mr{h,rms}}^{2}$ in Fig.~2c over the entire range of $|V_{\mr{tg}}-V_{\mr{D}}|$ confirms that the $\Delta T$ is solely proportional to the heating current and independent of the doping induced by the top gate. The independent scaling of $V_{2\omega,\mr{rms}}$ with $I_{\mr{h,rms}}^{2}$ and $\Delta T$ with $I_{\mr{h,rms}}^{2}$ (inset Fig.~2c) confirms that the all the measurements were done in the linear heating response regime and the condition $\Delta T\ll T$ is maintained throughout.

\subsection{Theoretical calculation.}

We have employed a phenomenological metal-insulator-metal junction model, where the interlayer tunneling occurs with tunneling probability $\Gamma(E)$. The Seebeck coefficient can then be written as,

\[
S=-\frac{1}{eT}\frac{\int_{-\infty}^{\infty}\mi{\Gamma}(E)D_{1}(E)D_{2}(E)(E-\mu)(-\frac{\partial f}{\partial E})dE}{\int_{-\infty}^{\infty}\mi{\Gamma}(E)D_{1}(E)D_{2}(E)(-\frac{\partial f}{\partial E})dE}
\]

\noindent where the density-of-states $D_{1,2}(E)$ are $\propto E$ for pristine graphene layers, and $f(E)$ is the Fermi function. In order to incorporate the effect of inhomogeneity at very low energies, we have parametrized the potential fluctuations with a broadening constant $\gamma$ as shown in the schematic of Fig.~3a, where $\gamma\approx55$~meV is estimated from the experimental transport parameters.

\section{acknowledgement}

The authors thank the financial support from the Department of Science and Technology, Government of India. P.S.M. and A.G. thank the National Nanofabrication Center, CeNSE, IISc (NNfC) for  clean room fabrication facilities and the Micro and Nano Characterization Facility, CeNSE, IISc (MNCF) for optical and mechanical characterization facilities.



\bibliographystyle{naturemag}
\bibliography{TBLG}


\end{document}